\documentclass[journal,twocolumn]{IEEEtran}
\usepackage{cmap}
\usepackage[T1]{fontenc}
\usepackage{amsmath,graphicx,hhline}
\usepackage{caption}
\usepackage{subcaption}
\usepackage{amssymb,amsthm}
\usepackage{mathrsfs}
\usepackage[nodisplayskipstretch]{setspace}
\usepackage{enumerate}
\usepackage{multirow}
\usepackage[update,prepend]{epstopdf}
\usepackage{float}
\restylefloat{table}
\usepackage{url}
\usepackage{listings}
\usepackage{algorithm}
\usepackage{algpseudocode}
\usepackage{booktabs}
\usepackage{lipsum}
\usepackage{geometry}

\newgeometry{
    top=1in,
    bottom=1in,
    outer=1in,
    inner=1in,
}

\newtheorem*{theorem*}{Theorem}
\newtheorem{theorem}{Theorem} 
\newtheorem*{problemstatement*}{Problem Statement}
\newtheorem*{definition*}{Definition}
\newtheorem{definition}[theorem]{Definition}
\newtheorem{result}[theorem]{Result}

\newtheorem*{Lemma*}{Lemma}
\newtheorem*{corollary*}{Corollary}

\newtheorem{property}[theorem]{Property}

\title{Agile Inexact Methods for Spectral Projector-Based Graph Fourier Transforms}
\author{Joya A. Deri,~\IEEEmembership{Member, IEEE,}
        and Jos\'{e} M. F. Moura,~\IEEEmembership{Fellow, IEEE}
        \thanks{This work was partially supported by NSF grants CCF-1011903 and CCF-1513936 and an SYS-CMU grant.}
        \thanks{The authors are with the Department of Electrical and Computer Engineering, Carnegie Mellon University, Pittsburgh, PA 15213 USA (email: {jderi,moura}@andrew.cmu.edu)}%
        }
\begin{document}
\maketitle
\begin{abstract}
We propose an inexact method for the graph Fourier transform of a graph signal, as defined by the signal decomposition over the Jordan subspaces of the graph adjacency matrix. This method projects the signal over the generalized eigenspaces of the adjacency matrix, which accelerates the transform computation over large, sparse, and directed adjacency matrices. The trade-off between execution time and fidelity to the original graph structure is discussed. In addition, properties such as a generalized Parseval's identity and total variation ordering of the generalized eigenspaces are discussed.

The method is applied to 2010-2013 NYC taxi trip data to  identify traffic hotspots on the Manhattan grid. Our results show that identical highly expressed geolocations can be identified with the inexact method and the method based on eigenvector projections, while reducing computation time by a factor of 26,000 and reducing energy dispersal among the spectral components corresponding to the multiple zero eigenvalue.
\end{abstract}
\begin{IEEEkeywords}
Jordan decomposition, generalized eigenspaces, directed graphs, sparse matrices, signal processing on graphs, large networks, inexact graph Fourier transforms
\end{IEEEkeywords}
\section{Introduction}
\label{sec:intro}
%
Graph signal processing~\cite{sandryhaila2013discrete,shuman2013emerging,teke2017extending,zhu2012approximating,narang2012perfect} allows for analysis of signals over graph structures with the framework of digital signal processing. Properties that have recently been explored include graph filtering~\cite{teke2017extending,narang2012perfect} and sampling and recovery of graph signals~\cite{marques2016sampling,segarra2016reconstruction,chen2015signal}. This paper focuses on considerations for applying adjacency matrix-based graph signal processing as in~\cite{sandryhaila2013discrete,sandryhaila2014big,sandryhaila2014discrete,deriGFT2016} to real-world networks. 

Real-world networks  are often characterized by two salient features:  (1) extremely large data sets and (2) non-ideal network properties that require extensive computation to obtain suitable descriptions of the network.  These features can result in extremely long execution times for the analyses of interest.  This paper presents methods that significantly reduce these computation times.

The adjacency matrices of real-world large, directed, and sparse networks may be \emph{defective}. For such matrices,~\cite{sandryhaila2013discrete} defines the eigendecomposition of the matrix in terms of the Jordan decomposition. This requires determination of Jordan chains, which are expensive to compute, potentially numerically unstable, and non-unique with respect to a set of fixed eigenvectors, resulting in non-unique coordinate graph Fourier transforms. In~\cite{deriGFT2016}, we reformulate the graph signal processing framework so that the graph Fourier transform is unique over defective networks. 

Since our objective is to provide methods that can be applied to real-world systems, it is necessary to consider the associated computational cost. Efficient algorithms and fast computation methods are essential, and with modern computing systems, parallelization and vectorization of software allow decreased computation times~\cite{franchetti2009}.   In this paper, we provide a method to accelerate computation by relaxing the determination of the Jordan decomposition of an adjacency matrix.

Consider a graph~$\mathcal G=G(A)$ with adjacency matrix $A\in\mathbb C^{N\times N}$ and a graph signal $s\in\mathbb C^N$, which characterizes properties or behaviors at each node of the graph. The graph Fourier transform of~\cite{sandryhaila2013discrete} is defined as follows. If the adjacency matrix has Jordan decomposition $A=VJV^{-1}$,  the graph Fourier transform of a signal $s\in\mathbb C^{N}$ over~$\mathcal G$ is defined as 
\begin{equation}
\label{eq:origgft}
\widetilde s = V^{-1}s,
\end{equation}
where $V^{-1}$ is the \emph{Fourier transform matrix} of $\mathcal G$. This is essentially a projection of the signal onto the proper and generalized eigenvectors of $A$. It is an orthogonal projection when $A$ is normal ($A^H A= A A^H$) and the eigenvectors form a unitary basis (i.e., $V^{-1}= V^H$).

For \emph{defective}, or non-diagonalizable adjacency matrices, which have at least one eigenvalue~$\lambda_i$ with algebraic multiplicity (the exponent in the characteristic polynomial of $A$) greater than the geometric multiplicity (the kernel dimension of $A-\lambda_i I$, or $\mathrm{dim}\,\mathrm{Ker}(A-\lambda_i I)$, where $I$ is the $N\times N$ identity matrix), the corresponding eigenvectors may not span $\mathbb C^N$. 
The basis can be completed by computing Jordan chains of generalized eigenvectors~\cite{lancaster1985,golub2013matrix}, but the computation introduces degrees of freedom that render these generalized eigenvectors non-unique. This is addressed in~\cite{deriGFT2016} by defining a spectral projector-based GFT in terms of projections onto the Jordan subspaces of the adjacency matrix of the graph. Formally, the GFT of a signal $s\in\mathbb C^N$ over graph $\mathcal G = G(A)$, where adjacency matrix $A$ with $k$ distinct eigenvalues has Jordan decomposition $A=VJV^{-1}$ and Jordan subspaces $\{\mathscr J_{ij}\}$, $i=1,\dots,k$, $j= 1,\dots, \mathrm{dim}\,\mathrm{Ker}(A-\lambda_i I)$, is defined as the mapping
\begin{align}
\mathcal F: \mathcal S &\rightarrow  \bigoplus_{i=1}^k \bigoplus_{j=1}^{g_i} \mathscr J_{ij}\nonumber\\
s&\rightarrow \left(\widehat s_{11},\dots,\widehat s_{1g_1},\dots,\widehat s_{k1},\dots,\widehat s_{kg_k}\right) \label{eq:gft},
\end{align}
where each $\widehat s_{ij}$ represents the projection onto Jordan subspace $\mathscr J_{ij}$ parallel to $\mathcal S\backslash \mathscr J_{ij}$.


The spectral projector GFT requires Jordan chain vector computations for non-diagonalizable, or \emph{defective} adjacency matrices, which can be exceedingly inefficient for large, directed matrices with many nontrivial Jordan subspaces. The desire for a more efficient transform computation motivates the inexact approaches proposed in this paper.
In particular, we propose an inexact method for computing the graph Fourier transform that simplifies the choice of projection subspaces. This method is motivated by insights based on  graph equivalence classes in~\cite{deriEquiv2016} and dramatically reduces computation on real-world networks.  The runtime vs. fidelity trade-off associated with this method is explored.

To obtain the inexact method, the spectral components are redefined as the generalized eigenspaces of adjacency matrix $A\in\mathbb C^{N\times N}$, and the resulting properties are discussed in Section~\ref{sec:AIM}. Section~\ref{sec:AIM:parseval} establishes the generalized Parseval's identity with respect to the inexact method. An equivalence class with respect to the inexact method is discussed in Section~\ref{sec:AIM:graphequiv}, which  leads to a total variation ordering of the new spectral components as discussed in Section~\ref{sec:AIM:totalvar}. Section~\ref{sec:AIM:reallworld} discusses the utility of the inexact method for real-world network analysis, and Section~\ref{sec:AIM:tradeoffs} describes the trade-offs associated with its use. 

Lastly, Section~\ref{sec:taxires} demonstrates the application of the inexact GFT to New York City taxi data on the Manhattan street network.  Our results demonstrate the speed of the inexact method and show that it reduces energy dispersal over spectral component corresponding to eigenvalue $\lambda =0$.
%
%
%
%



%
%
%
%
\section{Agile Inexact Method for Accelerating Graph Fourier Transform Computations}
\label{sec:AIM}
This section defines the Agile Inexact Method (AIM) for computing the graph Fourier transform~\eqref{eq:gft} (GFT) of a signal. The \emph{inexactness} of the AIM results from the abstraction of the Jordan subspaces that characterize~\eqref{eq:gft}; this concept is  discussed further in this section. The \emph{agility} of the AIM is related to the Jordan equivalence classes introduced in~\cite{deriEquiv2016}, which demonstrate that the degrees of freedom allowed by defective matrices yield GFT equivalence across networks of various topologies. Sections~\ref{sec:AIM:graphequiv},~\ref{sec:AIM:reallworld}, and~\ref{sec:AIM:tradeoffs} discuss the increased computational efficiency that the AIM allows. 
We emphasize that reducing computation time is a major objective for formulating the inexact method.

The Agile Inexact Method (AIM) for computing the graph Fourier transform of a signal is defined as the signal projections onto the generalized eigenspaces of the adjacency matrix~$A$. When $A$ has $k$ distinct eigenvalues, the definition of a generalized eigenspace~$\mathscr G_i$ (see~\cite{lancaster1985,horn2012matrix}) is
\begin{equation}
\label{eq:generalizedeigenspace}
\mathscr G_i = \mathrm{Ker}\left(A - \lambda_i I\right)^{m_i},\,i=1\dots,k
\end{equation}
for corresponding eigenvalue $\lambda_i$ and eigenvalue index~$m_i$. Each $\mathscr G_i$ is the direct sum of the Jordan subspaces of $\lambda_i$. 

The direct sum of generalized eigenspaces provides a unique decomposition of the signal space. This leads to the definition of the \emph{Agile Inexact Method} for computing the graph Fourier transform of a graph signal $s\in\mathcal S$
\begin{align}
\mathcal H: \mathcal S &\rightarrow  \bigoplus_{i=1}^k \mathscr G_{i}\nonumber\\
s&\rightarrow \left(\breve s_{1},\dots,\breve s_{k}\right) \label{eq:AIMgft}.
\end{align}
The method when applied to signal~$s$ yields the unique decomposition
\begin{equation}
\label{eq:AIMgftsum}
s = \sum_{i=1}^k \breve s_{i},\hspace{1cm}\breve s_{i}\in \mathscr G_{i},
\end{equation}
where each $\breve s_i$ is the (oblique) projection of $s$ onto the $i$th generalized eigenspace $\mathscr G_i$ parallel to $\mathcal S \backslash \mathscr G_i$, or  $\bigoplus_{j\neq i}\mathscr G_j$.

We define the projection operator for the inexact method. Denote eigenvector matrix 
\begin{equation}
V=[v_{1,1}\cdots v_{1,a_1}\cdots v_{k,1}\cdots v_{k,a_k} ],
\end{equation}
where $v_{i,j}$ represents an eigenvector or generalized eigenvector of $\lambda_i$ and $a_i$ is the algebraic multiplicity of~$\lambda_i$, $i=1,\dots,k$, $j=1,\dots,a_i$. Let $V^0_i$ represent the $N\times N$ matrix that consists of zero entries except for columns containing $v_{i,1}$ through $v_{i,a_i}$. The signal expansion components for signal $s$ are $\widetilde s = Vs$ for signal $s\in\mathbb C^N$, where 
\begin{equation}
\widetilde s = [\widetilde s_{1,1}\cdots \widetilde s_{1,a_1}\cdots \widetilde s_{k,1}\cdots \widetilde s_{k,a_k}]^T.
\end{equation}
Therefore,
\begin{align}
\breve s_i &= \widetilde s_{i,1}v_{i,1}+\cdots\widetilde s_{i,a_i}v_{i,a_i}\\
&=V^0_i \widetilde s\\
&=V^0_i V^{-1}s\\
&= V \begin{bmatrix}
\ddots \\& I_{a_i} \\ && \ddots
\end{bmatrix} V^{-1}s\label{eq:AIMprojection_matrixall}\\
&= Z_{i0} s,
\end{align}
where \begin{equation}\label{eq:Zi0}Z_{i0}=V \begin{bmatrix}
\ddots \\& I_{a_i} \\ && \ddots
\end{bmatrix} V^{-1}\end{equation} is the \emph{first component matrix} of the $i$th generalized eigenspace~\cite{lancaster1985}. Missing entries in~\eqref{eq:AIMprojection_matrixall} and~\eqref{eq:Zi0}, including dot entries, are zero. The matrix $Z_{i0}$~\eqref{eq:Zi0} projects $s\in\mathbb C^N$ onto the generalized eigenspace~$\mathscr G_i$ parallel to $\bigoplus_{j\neq i}\mathscr G_j$. Note that the projection may be oblique. The reader is directed to  Theorems~9.5.2 and~9.5.4 in~\cite{lancaster1985} for more properties of~\eqref{eq:Zi0}.

The AIM provides a unique, Fourier transform-like projection space for a signal that is analogous to the definition~\eqref{eq:gft} based on Jordan subspaces because the signal space is a direct sum of  the generalized eigenspaces. 
The generalized eigenspaces are not necessarily irreducible components of the signal space, however, while the Jordan subspaces do provide an irreducible decomposition of the signal space. In terms of algebraic signal processing~\cite{puschel2008algebraic_foundation,puschel2008algebraic_1dspace}, this means that the Jordan subspace formulation~\eqref{eq:gft} can define a Fourier transform of a signal, while the generalized subspace~\eqref{eq:AIMgft} is inexact, and is not strictly a formulation of a true graph Fourier transform. The projections onto the generalized eigenspaces are additions of the spectral components of the GFT~\eqref{eq:gft} and can be considered to be inexact analogues to the spectral components. 

Note that the AIM~\eqref{eq:AIMgft} resolves to the Jordan subspace-based GFT~\eqref{eq:gft} whenever the maximum number of Jordan subspaces per distinct eigenvalue is one (i.e., the characteristic and minimal polynomials of~$A$ are equal). Otherwise, Jordan subspace information is lost when the AIM is applied.

The following sections present key properties of~\eqref{eq:AIMgft}, with particular emphasis on the generalized Parseval's identity, graph equivalence classes with respect to~\eqref{eq:AIMgft}, and total variation ordering.
\section{Generalized Parseval's Identity}
\label{sec:AIM:parseval}
Since the full graph Fourier basis~$V$ may not be unitary, the signal energy can be characterized by a generalized Parseval's identity. This identity is defined in~\cite{jelena2014foundations} and used in~\cite{deriGFT2016} to characterize the spectral components. The identity is presented as a property below:
\begin{property}[Generalized Parseval's Identity]
\label{prop:generalizedparseval}
Consider graph signal
~$s\in\mathbb C^N$ over graph $\mathcal G(A)$,~$A\in\mathbb C^{N\times N}$. Let $V = [v_1 \cdots v_N]$ be a Jordan basis for~$A$ with dual basis matrix $W = V^{-H}$ partitioned as $[w_1 \cdots w_N]$. Let $s = \sum_{i=1}^N \langle s,v_i \rangle v_i = V\widetilde s_V$ be the representation of $s$ in basis~$V$ and $s = \sum_{i=1}^N \langle s,w_i \rangle w_i = W\widetilde s_W$ be the representation of~$s$ in basis~$W$. Then
\begin{equation}
\label{eq:parseval_equalargs}
\left\| s\right\|^2 = \langle s,s  \rangle = \langle \widetilde s_{V},\widetilde s_{W} \rangle. 
\end{equation}
\end{property}
In this way, a biorthogonal basis set can be used to define signal energy. In particular, for a signal $s$ over graph $\mathcal G = \mathcal G(A)$, $V$ represents the eigenvector matrix of graph adjacency matrix~$A$, and the columns of $V$ and $W= V^{-H}$ form a dual basis. Thus, the expansions of signal~$s$ in these bases yield coefficient vectors~$\widetilde s_V$ and~$\widetilde s_W$ that satisfy~\eqref{eq:parseval_equalargs}.

We define the energy of a signal projection onto an AIM GFT component~$\breve s_i$. Suppose $\breve s_i = \mathrm{span}(v_{i,1},\dots,v_{i,a_i})$, where $v_{i,1},\dots,v_{i,a_i}$ are columns of $V$ with corresponding columns $w_{i,1},\dots,w_{i,a_i}$ of~$W$.  For signal $s$ over graph $\mathcal G = \mathcal G(A)$, write~$\breve s_i$ in terms of the columns of $V$ as \begin{equation}\breve s_i = \alpha_1 v_{i,1}+\dots +\alpha_{a_i}v_{i,a_i}\end{equation} and in terms of the columns of $W$ as  \begin{equation}\breve s_i = \beta_1 w_{i,1}+\dots \beta_{a_i}w_{i,a_i}.\end{equation} Note that $\alpha = (\alpha_1,\dots,\alpha_{a_i})$ and $\beta = (\beta_1,\dots,\beta_{a_i})$ are subsets of coefficient vectors $\widetilde s_V$ and $\widetilde s_W$ that satisfy~\eqref{eq:parseval_equalargs}, respectively. The energy of~$\breve s_i$ can then be defined as
\begin{equation}
\label{eq:AIMgftenergy}
\|\breve s_i\|^2 = \langle \alpha,\beta\rangle.
\end{equation}
Equation~\eqref{eq:AIMgftenergy} is used in Section~\ref{sec:taxires} to compare the inexact method~\eqref{eq:AIMgft} and original GFT~\eqref{eq:gft}.
\section{Graph Equivalence under the Agile Inexact Method}
\label{sec:AIM:graphequiv}
Just as different choices of Jordan subspace bases yield the same GFT over Jordan equivalence classes of graph topologies in~\cite{deriGFT2016}, different choices of bases for the generalized eigenspaces of the adjacency matrix yield the same inexact GFT. We define a $\mathscr G$-equivalence class as follows:
\begin{definition}[$\mathscr G$-Equivalent Graphs]
\label{def:genequivgraph}
Consider graphs~$\mathcal G(A)$ and~$ \mathcal G(B)$ with adjacency matrices $A,B\in\mathbb C^{N\times N}$. Then $\mathcal G(A)$ and $\mathcal G(B)$ are \emph{$\mathscr G$-equivalent graphs} if all of the following are true:
\begin{enumerate}
\item The eigenvalues  of $A$ and $B$ are identical (including algebraic and geometric multiplicities) -- that is, $\mathcal G_A$ and $\mathcal G_B$ are {cospectral}; and
\item $\{\mathscr G_{A,i}\}_{i=1}^k = \{\mathscr G_{B,i}\}_{i=1}^k$, where $\mathscr G_{A,i}$ and $\mathscr G_{B_i}$ are the $i$th generalized eigenspaces of $A$ and $B$, respectively, and  $k$ is the number of distinct eigenvalues.
\end{enumerate} 
\end{definition}
The set of all graphs that satisfy Definition~\ref{def:genequivgraph} with respect to a graph adjacency matrix $A$ with Jordan decomposition $A=VJV^{-1}$ form the $\mathscr G$-equivalence class of $\mathcal G(A)$. Denote by $\mathbf G_A$ the $\mathscr G$-equivalence class of~$\mathcal G(A)$. 

The power of the $\mathscr G$-equivalence classes comes from the observation that  a graph adjacency matrix $A= VJV^{-1}$ may have the same eigenvalues and generalized eigenspaces as  a matrix $B= \widetilde{V} J \widetilde V^{-1}$. For the AIM method~\eqref{eq:AIMgft}, the basis provided by the columns of $\widetilde V$ yields identical signal projections as that provided by the columns of~$V$. In addition, when $\widetilde V$ can be computed without finding Jordan chain vectors, the overall computation time for the eigendecomposition decreases.  Section~\ref{sec:AIM:tradeoffs} demonstrates the utility of these equivalence classes.

In contrast to the Jordan equivalence classes in~\cite{deriEquiv2016}, $\mathscr G$-equivalence classes are independent of Jordan subspaces and their dimensions
. This allows for greater degrees of freedom in basis choices that yield the same signal projections. 
As a result, the eigendecomposition required for the AIM~\eqref{eq:AIMgft} can be simpler and faster to compute than that for the GFT~\eqref{eq:gft}. A particular case is discussed in greater detail in Section~\ref{sec:AIM:tradeoffs}. 

The next section defines a total variation ordering over the generalized eigenspaces~$\mathscr G_i$~\eqref{eq:generalizedeigenspace} of the graph adjacency matrix. A class total variation is proposed to account for different choices of Fourier bases that correspond to a particular $\mathscr G$-equivalence class.

\section{Total Variation Ordering}
\label{sec:AIM:totalvar}
In~\cite{deriGFT2016} and~\cite{deriEquiv2016}, we defined an ordering on the GFT spectral components based on the total variation of a signal. This  allows ordering the frequency components so as to define low-pass, band-pass, and high-pass graph signals with respect to specific graphs. Using a total variation-based characterization signifies that the variation between a signal and its transformation by the graph adjacency matrix is less for low-pass signals than for high-pass signals.   In this section, such an ordering is shown for GFT spectral components corresponding to the Agile Inexact Method (AIM).

Denote by $V_i$ the submatrix of columns of eigenvector matrix $V$ that span $\mathscr G_i$, the $i$th generalized eigenspace. The (graph) total variation of $V_{i}$ is defined as
\begin{equation}
\label{eq:TV_singlecomp_AIM}
\mathrm{TV}_G\left(V_{i}\right) =  \left\|V_{i} - A V_{i} \right\|_1.
\end{equation}

The definition~\eqref{eq:TV_singlecomp_AIM} of total variation has been characterized in~\cite{deriGFT2016} where each $V_i$ spans a Jordan subspace instead of a generalized eigenspace. Since the generalized eigenspaces are direct sums of Jordan subspaces (see Section~\ref{sec:intro}), many properties hold for the total variation over generalized eigenspaces. Theorem~\ref{thm:TVisomorphic_AIM} shows that the total variation is invariant to a permutation of node labels. Theorem~\ref{thm:TV_ub_AIM} shows an upper bound on this total variation.  These theorems follow immediately from the proofs of Theorems~25 and~28 of~\cite{deriGFT2016}. These theorems show the dependence of the total variation~\eqref{eq:TV_singlecomp_AIM} on the choice of eigenvector matrix $V$ and motivate the definition of basis-independent class total variations as in~\cite{deriEquiv2016}. 
\begin{theorem}
\label{thm:TVisomorphic_AIM}
Let $A,B\in\mathbb C^{N\times N}$ such that $\mathcal G(B)$ is isomorphic to $\mathcal G(A)$. Let $V_{A,i}\in\mathbb C^{N\times_{a_{i}}}$ be a union of Jordan chains that span generalized eigenspace~$\mathscr G_i$ of matrix $A$ and $V_{B,i}\in\mathbb C^{N\times_{a_{i}}}$ the corresponding union of Jordan chains of $B$. Then $\mathrm{TV}_G(V_{A,i}) = \mathrm{TV}_G(V_{B,i})$.
\end{theorem}
We briefly note that the generalized eigenspaces of the adjacency matrices $A$ and $B$ in Theorem~\ref{thm:TVisomorphic_AIM} are not necessarily equivalent but are related by an isomorphism. The reader is referred to~\cite{deriEquiv2016} for more on isomorphic equivalence classes.
%
\begin{theorem}
\label{thm:TV_ub_AIM}
Consider adjacency matrix $A$ with $k$ distinct eigenvalues and $N\times a_{i}$ eigenvector submatrices~$V_{i}$ with columns corresponding to the union of the Jordan chains of $\lambda_i$, $i=1,\dots,k$. Then the graph total variation is bounded as \begin{equation}\label{eq:TVineq}\mathrm{TV}_G(V_{i})\leq \left|1-\lambda_i \right|+1.\end{equation}
\end{theorem}

The bound~\eqref{eq:TVineq} highlights that the choice of Jordan chain vectors in the eigendecomposition of the adjacency matrix affects the total variation. In order to gain independence from a choice of basis, the definition of total variation can be generalized to a class total variation defined over the $\mathscr G$-equivalence class associated with the graph of adjacency matrix~$A$. We define the \emph{class total variation} of spectral component $\mathscr G_{i}$ as the supremum of the graph total variation of $V_{i}$ over the $\mathscr G$-equivalence class (for all $\mathcal G(B) \in\mathbf G_A$):
\begin{align}
\label{eq:TV_H}
\mathrm{TV}_{\mathbf G_A}\left(\mathscr G_{i}\right) & = \sup_{\substack{\mathcal G\left(B\right)\in\mathbf G_A\\ B = VJV^{-1}\\ \mathrm{span}\left\{V_{i}\right\} = \mathscr G_{i}\\ \left\| V_{i} \right\|_1 =1 }} \mathrm{TV}_G\left( V_{i}\right).
\end{align}

In particular, the low-to-high variation sorting is achieved via the function $f(\lambda_i) =\left|1-\lambda_i\right| +1$. 
\section{Applicability to Real-World Networks}
\label{sec:AIM:reallworld}
This section discusses how the AIM can be applied on real-world large, sparse, and directed networks. We discuss examples that illustrate that adjacency matrices for these networks  have a single generalized eigenspace of dimension greater than one. We demonstrate the ease of computing the AIM for such networks.
\subsection{Large, Directed, and Sparse Networks}
\label{sec:AIM:reallworld:sparse}
In practice, sparsity in directed networks yields a zero eigenvalue of high algebraic and geometric multiplicities that may not be equal. This behavior can be illustrated using two examples.  The first is a directed, strongly connected  road network~$\mathcal G_{\mathrm{rd}} = \mathcal G(A_\mathrm{rd})=(\mathcal V_\mathrm{rd},\mathcal E_\mathrm{rd})$ of Manhattan, New York~\cite{deriasilomar2015}; $\mathcal G_{\mathrm{rd}}$ contains~6,408 nodes that represent the latitude and longitude coordinates of intersections and mid-intersection locations (available from~\cite{data:roadnetwork}), as well as 14,418 directed edges, where edge~$e=(v_1,v_2)$ represents a road segment along which traffic is legally allowed to move from~$v_1$ to~$v_2$ as determined from Google Maps~\cite{deriasilomar2015}.  The second network example is the largest weakly connected component~$\mathcal G_{\mathrm{bl}} = \mathcal G(A_{\mathrm{bl}})=(\mathcal V_\mathrm{bl},\mathcal E_\mathrm{bl})$ of a political blog network, with~1,222 blogs as nodes and~19,024 directed edges, where edge~$e=(v_1,v_2)$ exists between blogs~$v_1,v_2\in \mathcal V_\mathrm{bl}$ if~$v_1$ contains a link to $v_2$~\cite{Adamic:2005}.

\begin{figure}[tb]
\captionsetup[subfigure]{justification=centering,singlelinecheck=on}
\begin{subfigure}[b]{0.25\textwidth}
\hspace{3.5cm}
\includegraphics[width=.75\linewidth]{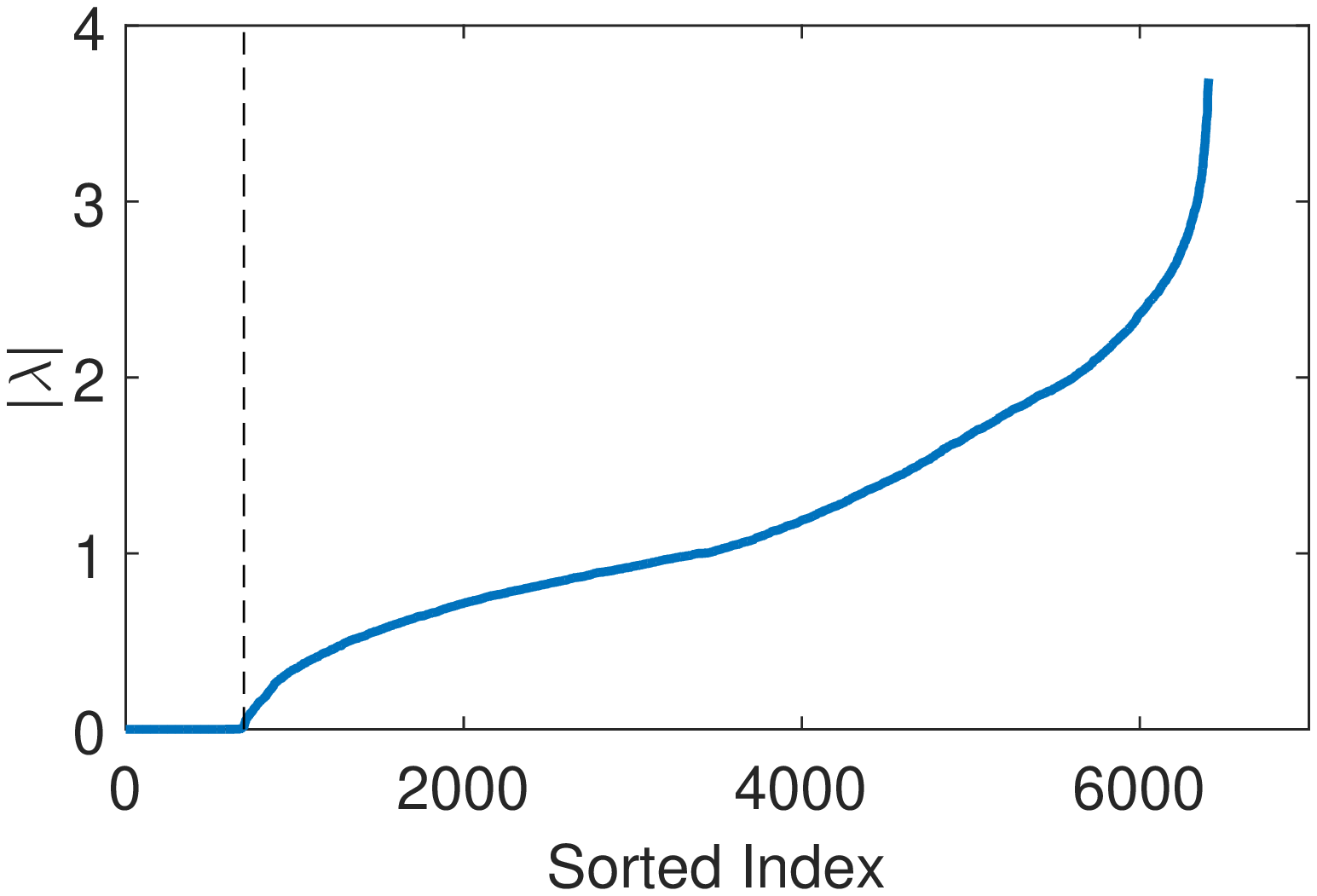}
\caption{Eigenvalue magnitudes\\ for road network} \label{fig:realworld_roads_abs_a}
\end{subfigure}
\hspace{-.38cm}
\begin{subfigure}[b]{0.25\textwidth}
\hspace{1.5cm}
\includegraphics[width=.7\linewidth]{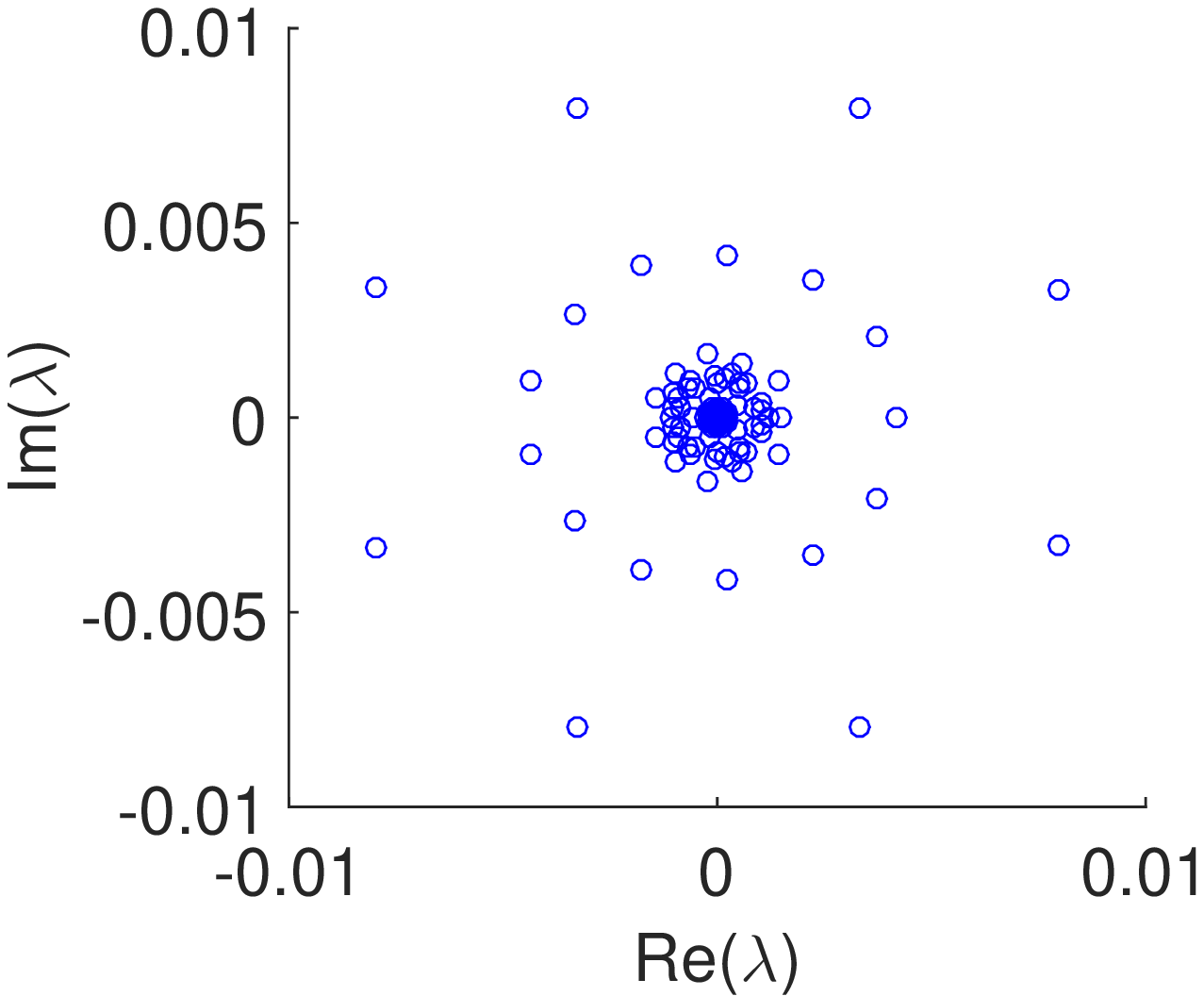}
\caption{Small eigenvalues\\ for road network} \label{fig:realworld_roads_complexplane_b}
\end{subfigure}
\begin{subfigure}[b]{0.25\textwidth}
\hspace{1.5cm}
\includegraphics[width=.75\linewidth]{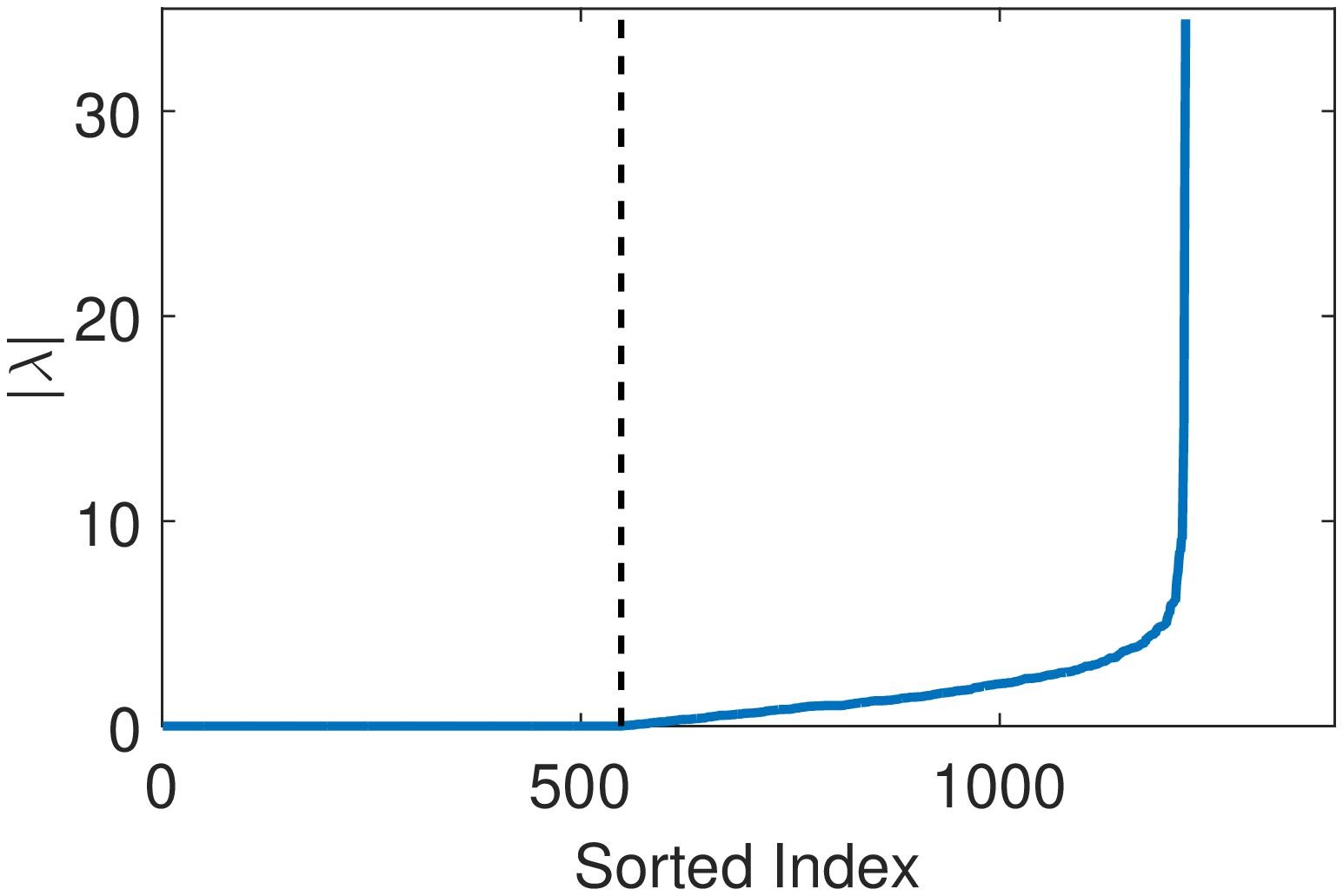}
\caption{Eigenvalue magnitudes\\ for political blog} \label{fig:realworld_polblog_abs_c}
\end{subfigure}
\hspace{-.38cm}
\begin{subfigure}[b]{0.25\textwidth}
\hspace{1.5cm}
\includegraphics[width=.7\linewidth]{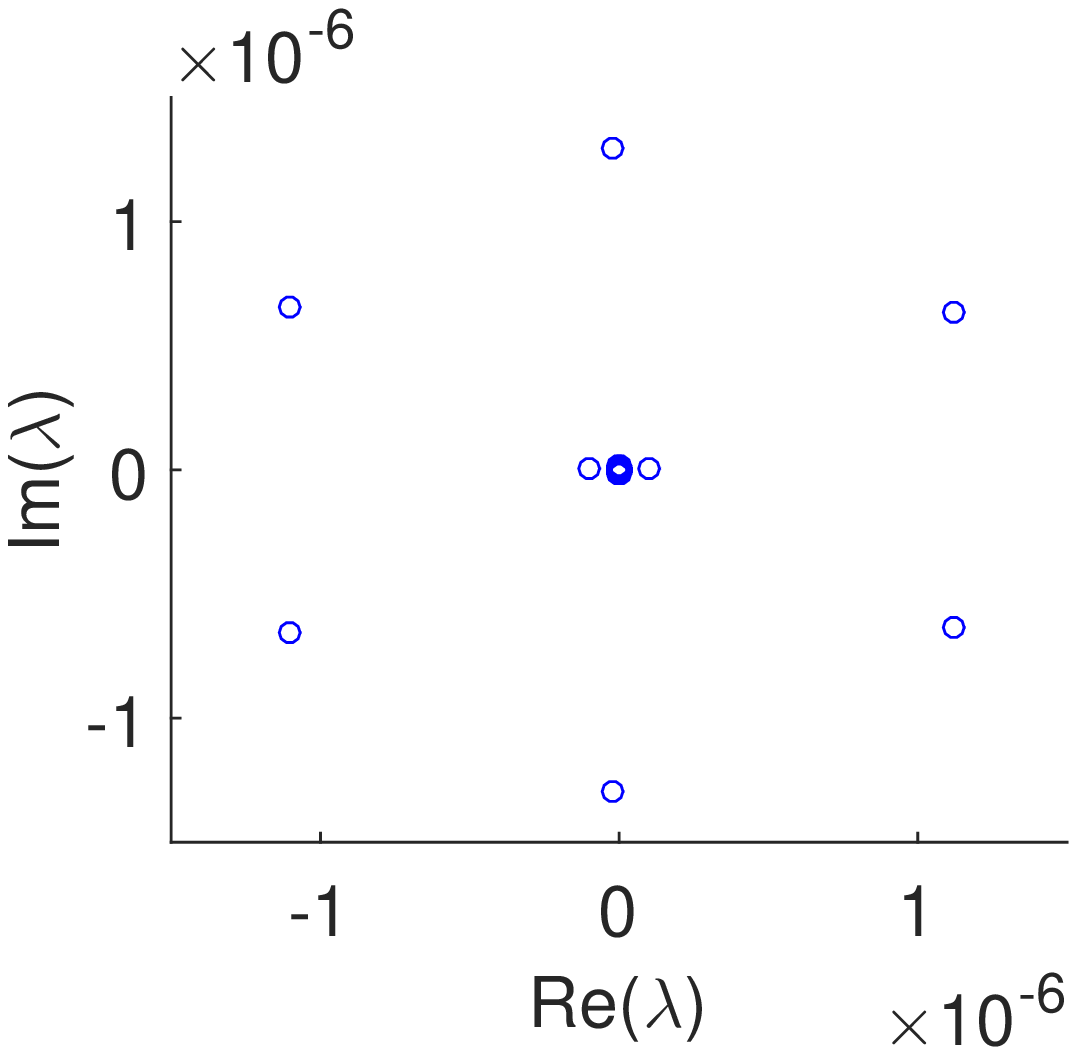}
\caption{Small eigenvalues\\ for political blog} \label{fig:realworld_polblog_complexplane_d}
\end{subfigure}
\caption{(a) Eigenvalue magnitudes of a directed New York City road network adjacency matrix. The magnitudes are small in the index range 1 to~699 (up to the dotted line). (b) Road network eigenvalues (699 total) in the sorted index range 1 to~699 plotted on the complex plane. (c) Eigenvalue magnitudes of the largest weakly connected component of a political blog network. The magnitudes are small in the index range 1 to 548 (up to the dotted line). (d) Blog eigenvalues (548 total) in the sorted index range 1 to 548.}
\label{fig:realworld}
\vspace{-.10cm}
\end{figure}
Figures~\ref{fig:realworld_roads_complexplane_b} and~\ref{fig:realworld_polblog_complexplane_d} show that the numerical eigenvalues of these real-world networks occur in clusters on the complex plane. These eigenvalues have small magnitude (lying left of the dashed lines in Figures~\ref{fig:realworld_roads_abs_a} and~\ref{fig:realworld_polblog_abs_c}). On first inspection, it is not obvious whether the numerical eigenvalues are spurious eigenvalues clustered around a true eigenvalue of zero, or whether the numerical eigenvalues are indeed the true eigenvalues.\footnote{The eigenvalues in Figure~\ref{fig:realworld} were computed in MATLAB. Similar clusters appear using \texttt{eig} without balancing and \texttt{schur}. They also appear when analyzing row-stochastic versions of the asymmetric adjacency matrices.} This phenomenon is typically observed in systems with multiple eigenvalues~\cite{kahan1972conserving,zeng_jordanformbkch,zeng_defeigenval}. In practice, we can verify that numerical zero is an eigenvalue using either pseudospectra~\cite{trefethen2005spectra} or techniques that explore the numerical stability of the singular value decomposition~\cite{ruhe1970,ruhe1970_veryill}. For the examples explored here, the singular value decomposition demonstrated that the clusters of low-magnitude eigenvalues represented a numerical zero eigenvalue. This method involves comparing small singular values to machine precision and is discussed in more detail in Section~\ref{sec:taxis:eigendecomp:verify0eigenval}.

Furthermore, the adjacency matrices~$A_{\mathrm{rd}}$ and~$A_{\mathrm{bl}}$  have kernels of dimension~$446$ and~$439$, respectively\footnote{The kernels (null spaces) were computed in MATLAB~\cite{tool:MATLAB:2015} on a 16-GB, 16-core Linux machine.}, confirming that  eigenvalue~$\lambda = 0$ has high algebraic (and geometric) multiplicity. 
In addition, eigenvector matrices~$V_{\mathrm{rd}}$ and~$V_{\mathrm{bl}}$ computed with 
MATLAB's eigensolver have numerical rank~$6363<6408$ and~$910<1222$, respectively, implying the existence of nontrivial Jordan subspaces  (Jordan chains of length greater than one) for~$\lambda = 0$. While these general eigenspace properties can be readily determined,  a substantial amount of additional computation is required to determine the dimensions of each Jordan subspace for eigenvalue~$\lambda = 0$. For example, the Jordan subspaces for~$\lambda=0$ can be deduced by computing a generalized null space as in~\cite{guglielminigeneralizednullspace2014}, but this computation takes $O(N^3)$ or $O(N^4)$ for an $N\times N$ matrix.

On the other hand, the eigenvectors corresponding to eigenvalues of higher magnitude in these networks are full rank.  In other words, from a numerical perspective, the Jordan subspaces for these eigenvalues are all one-dimensional. 

In such networks, for which sparsity yields a high-multiplicity zero eigenvalue while the other eigenvalues correspond to a full-rank eigenvector submatrix, the AIM inexact method given in~\eqref{eq:AIMgft} can be applied in the following way. Denote the eigenvector matrix~$V$ of graph adjacency matrix~$A$ such that
\begin{equation}
\label{eq:V_applyAIM}
V = \begin{bmatrix}
V_{\mathrm{known}} & \widetilde{V}
\end{bmatrix},
\end{equation}
where $V_{\mathrm{known}}$ be the full-rank eigenvector submatrix corresponding to the nonzero eigenvalues, and the columns of  submatrix $\widetilde V$ span the generalized eigenspace $\mathcal G_0$ corresponding to eigenvalue zero. 

One way to find a spanning set of columns for~$\widetilde V$  is to find the kernel of $V_{\mathrm{known}}^T$, since the range space of~
The resulting inexact graph Fourier basis is then
\begin{equation}
\label{eq:AIMV}
\widehat V = \begin{bmatrix}
V_{\mathrm{known}} & \mathrm{Ker}\left(V_{\mathrm{known}}^T\right)
\end{bmatrix}.
\end{equation}
If the original network has adjacency matrix $A=VJV^{-1}$, then~\eqref{eq:AIMV} implies that the network with adjacency matrix $\widehat A=\widehat V J\widehat V ^{-1}$ is in the same $\mathscr G$-equivalence class as the original network with adjacency matrix $A = VJV^{-1}$. As discussed in Section~\ref{sec:AIM:graphequiv},  the AIM GFT is equivalent over both networks. Section~\ref{sec:AIM:tradeoffs} shows that computing the Fourier basis as in~\eqref{eq:AIMV} is computationally fast and efficient.

In this way, for large, sparse, and directed real-world networks that exhibit a single zero eigenvalue of high multiplicity, the AIM can simplify the computation of a graph Fourier basis. Instead of computing Jordan chains or generalized null spaces to deduce the structure of the Jordan subspaces, a single kernel computation is needed to compute the spanning basis. This idea is explored further in Section~\ref{sec:AIM:tradeoffs}.
\subsection{Multiple Nontrivial Jordan Subspaces}
\label{sec:AIM:reallworld:multiplenontrivial}
In the case of multiple distinct eigenvalues with large but unequal algebraic and geometric multiplicities, it may be possible to compute a few Jordan chain vectors to differentiate the nontrivial generalized eigenspaces. This approach is only computationally efficient if the number of Jordan chain vectors to compute is relatively small, as discussed in~\cite{golub2013matrix,golub1976}.

When this is not feasible, a coarser inexact method can be obtained by reformulating~\eqref{eq:AIMV} so that $V_{\mathrm{known}}$ consists of all known eigenvectors and is full rank. Then a basis of the kernel of $V_{\mathrm{known}}^T$  can be computed to obtain a set of spanning vectors.  

It is unknown a priori which basis vectors of $\mathrm{Ker}\,V_{\mathrm{known}}^T$  belong to the nontrivial generalized eigenspaces of~$A$. This can be overcome by randomly assigning these vectors to the nontrivial generalized eigenspaces, which results in a coarser inexact method. The dimensions of each generalized eigenspace must be known so that the correct number of vectors is assigned to each eigenspace. 
The dimensions can be determined by recursively computing for eigenvalue $\lambda_i$
\begin{equation}
\label{eq:fl_diffkerdim}
f(l) = \mathrm{dim}\,\mathrm{Ker}(A-\lambda_iI)^l - \mathrm{dim}\,\mathrm{Ker}(A-\lambda_iI)^{l-1}
\end{equation}
 for $l=2,\dots,N$; this equation provides the number of Jordan chains of length at least~$l$~\cite{lancaster1985}. The value of $\mathrm{dim}\,\mathrm{Ker}(A-\lambda_iI)^{l-1}$ for~$l>1$ at which $f(l)=0$ is the  dimension of generalized eigenspace~$\mathscr G_i$. If $\left|\lambda_{\mathrm{max}}\right|>1$, the condition $f(l)=0$ may not be attained; instead,~$f(l)$ becomes a monotonically increasing function for large~$l$. In this case, the value of $\mathrm{dim}\,\mathrm{Ker}(A-\lambda_iI)^{l-1}$ for~$l$ at which~$f(l)$ becomes monotonically increasing is the approximate generalized eigenspace dimension. 

The random assignment of missing eigenvectors to the nontrivial generalized eigenspaces would be a coarser inexact method compared to the AIM method~\eqref{eq:AIMgft}; in particular, it is unknown which eigenvector assignment, if any, corresponds to a graph in the same $\mathscr G$-equivalence class as the original graph. On the other hand, the matrices built from a series of random assignments could be used to construct a filter bank such that each assignment corresponds to a different graph filter. An  optimal or near-optimal weighted combination of such filters  could be learned by using time-series of filtered signals as inputs to train a classifier. While the resulting Fourier basis is only an approximation of the one required for the AIM method~\eqref{eq:AIMgft}, such a learned combination of filters would provide an interpretable tool to analyze graph signals while also ranking each filter (representing a random assignment of vectors to generalized eigenspaces) in terms of its utility for typical graph signals.  

The next section characterizes the loss of fidelity to the original graph  when the AIM is applied as in~\eqref{eq:AIMV} for a single unknown, nontrivial generalized eigenspace.
\section{Runtime vs. Fidelity Trade-off}
\label{sec:AIM:tradeoffs}
While maximizing the fidelity of the network representation is desirable in general, obtaining as much knowledge of the Jordan chains as possible is computationally expensive.  As shown below, execution time is linear with respect to the number of Jordan chain vectors to compute. In practice, however, computation of increasing numbers of chains requires allocating memory for matrices of increasing size.  This memory allocation time can increase nonlinearly once specific limits (e.g., RAM) imposed by the computing hardware are exceeded.  Since practical applications of GFT analysis may be constrained by execution time requirements and available computing hardware, methods that allow fidelity vs. speedup tradeoffs will prove useful.

Our inexact GFT approach provides precisely such a method for trading between higher fidelity analysis (a greater number of Jordan chains computed) and execution time, as discussed in Section~\ref{sec:AIM:reallworld}.    In this section, the details of such trade-offs are discussed in detail.

To illustrate the cost of computing the Jordan chain vectors of $A\in\mathbb C^{N\times N}$ that complete the generalized eigenspace of eigenvalue~$\lambda_i$, consider the general algorithm given by:  

\vspace{1.5em}
\begin{spacing}{0.8}
\begin{algorithmic}[1]
\Function  {Compute\_Chains}{}
	\State V = \Call{Full\_Rank\_Eigenvectors}{$A$}
\State R = \Call{Pseudoinverse}{$A-\lambda_i I$}
\State N = \Call{Ker}{$A-\lambda_iI$}
\State Nnew=[\,\,\,]
\For{c in 1..maximum chain length}
\For{v in N}
\If {[$(A -\lambda_i I)$\,  v] full rank} 
\State v2 = R*v
\State Append v2 to V.
\State Append v2 to Nnew.
\EndIf
\EndFor
\State N = Nnew
\EndFor
\EndFunction
\end{algorithmic}
\end{spacing}
\vspace{1.5em}

The eigenvector and pseudoinverse computations are pre-processing steps that can be executed just once for a network with stationary topology. Efficient methods for these computations include power iterations, QR algorithms, and, for large, sparse matrices, Lanczos algorithms and Krylov subspace methods~\cite{golub2013matrix}.  The key bottlenecks of concern reside in the \texttt{for} loop. This loop consists of a rank-checking step to verify that the current eigenvector is in the range space of $A-\lambda_i I$. This step can be implemented using the singular value decomposition, which has $O(kMN^2 + k'M^3)$ operations~\cite{golub2013matrix} on an $M\times N$ matrix, $M>N$; constants $k$ and $k'$ could be 4 and 22, respectively, as in the R-SVD algorithm of~\cite{golub2013matrix}.  
In addition, the matrix-vector product in the \texttt{for} loop takes about $2MN$ operations~\cite{golub2013matrix}. The original dimension of~$V$ is $M\times N(j)$, where the number of columns $N(j)$ approaches $M$ from below as the number~$j$ of vectors traversed increases. The resulting time complexity for a single iteration is 
\begin{equation}
O(kMN(j)^2 + k'M^3) + O(2MN(j)).
\end{equation}

The first \texttt{for} loop in \texttt{compute\_chains} iterates~$m_i$ times, where~$m_i$ is the maximum chain length corresponding to $\lambda_i$. A method for estimating~$m_i$ is demonstrated in Section~\ref{sec:taxis:eigendecomp:jordanchaincompute}. The second \texttt{for} loop will first run for $g_i$ iterations, where $g_i$ is the kernel dimension of $A-\lambda_i I$ (the geometric multiplicity of $\lambda_i$). On the $l$th run of the outer loop, the number of iterations of the inner loop  equals the difference of kernel dimensions $f(l)$ given by~\eqref{eq:fl_diffkerdim}. 
Therefore, the total number~$b_A$ of iterations is $m_i \sum_{l=1}^{m_i}f(l)$, or
\begin{align}
&m_i(\mathrm{dim}\,\mathrm{Ker}(A-\lambda_iI)^{m_i} - \mathrm{dim}\,\mathrm{Ker}(A-\lambda_iI) )\\&=m_i(a_i - g_i ),\label{eq:totalbA}
\end{align}
 where $a_i$ and $g_i$ are the algebraic and geometric multiplicities of $\lambda_i$, respectively. Since $b_A$~\eqref{eq:totalbA} depends on the adjacency matrix~$A$, it has an implicit dependence on the $M\times N(j)$ dimensions of~$A$. The total time complexity of the \texttt{for} loops is then 
 \begin{equation}
 \sum_{j=1}^{b_A} O(kMN(j)^2 + k'M^3) + O(2MN(j)).
 \end{equation}

In addition, the time to allocate memory for the matrix {[$(A -\lambda I)$\,  v] is \begin{equation} O(MN(j))\cdot m(MN(j)),\end{equation} 
where $m(\cdot)$ is the platform-dependent time per unit of memory allocation as a function of the matrix size. Since each \texttt{for} loop allocates memory in this way, the total memory allocation time is \begin{equation}\sum_{j=1}^{b_A} O(MN(j))\cdot m(MN(j)).\end{equation} 
Assuming~$c$ is the platform-dependent time per floating-point operation and SVD constants $k=4$ and $k'=22$, the total expected runtime for the Jordan chain computation can be approximated as
\begin{align}
\sum_{j=1}^{b_A} c( 4MN(j)^2 &+ 22M^3 + 2MN(j))\nonumber\\
&+ MN(j)  m\left(MN(j)\right),\label{eq:runtime_jordanchain}
\end{align}
where $c$ is the platform-dependent time per floating-point operation and the full SVD complexity coefficients are set to~$k=4$ and~$k'=22$.

Equation~\eqref{eq:runtime_jordanchain} 
shows that the runtime for the Jordan chain computations is approximately linear with the number of missing vectors one needs to compute. In practice, however, the time to allocate memory can scale nonlinearly with the number of nodes as the size of the eigenvector matrix approaches system capacity. For networks at massive scale, this can present a significant problem.

In contrast, the algorithm to compute an orthogonal set of missing vectors is very fast. The algorithm is the following:

\vspace{1.5em}
\begin{spacing}{0.8}
\begin{algorithmic}[1]
\Function {compute\_missing}{}
\State V = \Call{Full\_Rank\_Eigenvectors}{$A$}
\State Vt = \Call{Transpose}{$A$}
\State Vnew = \Call{Ker}{Vt}
\State V = [V\,\, Vnew]
\EndFunction
\end{algorithmic}
\end{spacing}
\vspace{1.5em}

The bottleneck in \texttt{compute\_missing} is computing the kernel of $V^T$, which has time complexity $O(kMN(j)^2 + k'M^3)$ for a SVD-based implementation. In addition, memory allocation for~Vt and~V each takes $O(MN(j))$ time. Notably, this allocation is only performed once in the AIM implementation~\eqref{eq:AIMV} ($O(1)$ in space), as compared to the $m_i$ allocations required in the \texttt{for} loop of \texttt{compute\_chains}. 

While the output~$V$ matrix in \texttt{compute\_missing} yields the same AIM results given by~\eqref{eq:AIMgft} as the original eigenvector matrix, the corresponding adjacency matrix~$\widetilde A$ may not preserve key structures in the original graph adjacency matrix~$A$. In other words, the trade-off for the faster computation time in \texttt{compute\_missing} is a loss of fidelity to the original graph. In order to improve the fidelity, one may compute a certain number~$b<b_A$ of Jordan chains as determined by RAM and other system constraints. In this way, the AIM can be used to selectively trade off execution time~\eqref{eq:runtime_jordanchain} against fidelity in the computations.

This section shows that the AIM can be employed to enable GFT analysis to meet execution time constraints by trading runtime for fidelity. 
The analysis of algorithms \texttt{compute\_chains} and \texttt{compute\_missing} quantifies the expected execution time in terms of a given number of Jordan chain vectors.  These considerations need to be addressed when applying the graph Fourier transform~\eqref{eq:gft} and the AIM method~\eqref{eq:AIMgft} to real-world problems. Such an application is shown in the next section. 
%
%
%
\section{New York City Taxi Data: Eigendecomposition and Graph Fourier Transform}    
\label{sec:taxires}
This section expands on the issues described in Section~\ref{sec:AIM:reallworld}  and presents details for computing the eigenvector matrix $V$ for the non-diagonalizable adjacency matrix~$A$ of the Manhattan road network. 

The Agile Inexact Method (AIM) for the graph Fourier transform developed in Section~\ref{sec:AIM} is then applied to the statistics computed from four years of New York City taxi data. Our method provides a fine-grained analysis of city traffic that should be useful for urban planners, such as in improving emergency vehicle routing.

Sections~\ref{sec:taxis:eigendecomp}--~\ref{sec:taxis:eigendecomp:generalizedeigenspacecompute} describe our method to find the eigendecomposition of the Manhattan road network. Section~\ref{sec:taxis:AIM} demonstrates the AIM for computing the graph Fourier transform. 
\subsection{Data Set Descriptions}
\label{sec:taxis:datasets}
Our goal is to extract behaviors over space and time that characterize taxi movement through New York City based on four years (2010-2013) of New York City taxi data~\cite{data:taxidata}.  Since the path of each taxi trip is unknown, an additional processing step is required to estimate taxi trajectories before extracting statistics of interest. For example, if a taxi picks up passengers at Times Square and drops them off at the Rockefeller Center, it is desirable to have statistics that capture not just trip data at the landmarks, but also at the intermediate locations. 

Estimating tax trajectories requires overlaying the taxi data on the New York City road network. Details on the taxi data and the road network are described as follows.
\subsubsection{NYC Taxi Data}
The 2010-2013 taxi data we work with consists of 700 million trips for 13,000 medallions~\cite{data:taxidata}. Each trip has about 20 descriptors including pick up and drop off timestamps, latitude, and longitude, as well as the passenger count, trip duration, trip distance, fare, tip, and tax paid. The data is available as~16.7~GB of compressed CSV files.

\textbf{Dynamic representation.} Since the available data  provides only static (start and end) information for each trip, an estimate of the taxi trajectory is needed in order to capture the taxi locations interspersed throughout the city at a given time slice.  Our method for estimating these trajectories, which is based on Dijkstra's algorithm, is described in detail in~\cite{deriFranzMoura2016dijkstra}. 

Once taxi paths are estimated, they are used to extract traffic behavior. The statistic of interest we consider here  is the average number of trips that pass through a given location at a given time of day.  

\subsubsection{NYC and Manhattan Road Networks}
The road network~$G=(V,E)$ consists of a set~$V$ of $\left|V\right| =79,234$ nodes and a set~$E$ of $\left|E\right|=223,966$ edges which we represent as a $\left|V\right|\times \left|V\right|$  adjacency matrix~$A$. The nodes in $V$ represent intersections and points along a road based on geo-data from~\cite{data:roadnetwork}. Each edge~$(v_i,v_j)\in E$, $v_i,v_j\in V$, corresponds to a road segment on which traffic may flow from geo-location $v_i$ to geo-location $v_j$ as determined by Google Maps~\cite{data:googlemapsnyc}. An edge of length $d_{ij}>0$ is represented by a nonzero entry in the adjacency matrix so that $[A]_{ij}=d_{ij}$.  A zero entry corresponds to the absence of an edge.

The network~$G$ is \emph{directed} since the allowed traffic directions along a road segment may be asymmetric. In addition,~$G$ is \emph{strongly connected} since a path (trajectory) exists from any geo-location to any other geo-location. 

Our analysis focuses on the Manhattan grid, which is a subgraph of $G$ with $6,408$ nodes and $14,418$ edges. This subgraph is also strongly connected. The eigendecomposition in Section~\ref{sec:taxis:eigendecomp} is based on this network. Expressing the average number of trips that pass through a given location at a fixed time of day as a vector over the Manhattan grid defines a graph signal to which the GFT~\eqref{eq:gft} and AIM method~\eqref{eq:AIMgft} can be applied.

\subsection{Eigendecomposition of Manhattan Road Network}
\label{sec:taxis:eigendecomp}
The Manhattan road network described in Section~\ref{sec:taxis:datasets} provides the adjacency matrix~$A\in\mathbb R^{6408\times 6408}$ for which we compute the eigendecomposition. For this particular adjacency matrix, the eigenvector matrix~$V_\mathrm{obs}\in\mathbb C^{6408\times 6408}$ generated by a standard eigenvalue solver such as MATLAB or LAPACK is not full rank, where the rank can be computed with either the singular value decomposition or the QR decomposition; i.e., $A$ is not diagonalizable.

In order to apply the GFT~\eqref{eq:gft} or the original projections onto each eigenvector as in~\cite{sandryhaila2013discrete}, Jordan chain computations are required. These computations are costly  as described in Section~\ref{sec:AIM:tradeoffs} and also potentially numerically unstable. In contrast, the AIM~\eqref{eq:AIMgft} does not require this step. 

Since it is instructive to compare the GFT with the AIM results, it is necessary to compute the Jordan chains here.  For this reason, the following sections detail our method for computing the Jordan chains for the Manhattan road network.
\begin{center}
\begin{table*}[t]
\centering
\renewcommand{\arraystretch}{1.3}
  \begin{tabular}{@{}lllll@{}}
    \toprule
	$k$ & $m_k$ &$N-m_k$ &$\sigma_{N-m_k}\left(A^k\right)$ & $\sigma_{N-m_k+1}\left(A^k\right)$ \\
	\midrule
	1&$446$&$5962$&$1.9270\times 10^{-3}$ &$1.2336\times 10^{-15}$\\
	2&$596$&$5812$&$2.1765\times 10^{-6}$ &$6.9633\times 10^{-16}$\\
	3&$654$&$5754$&$1.4013\times 10^{-8}$ &$3.4250\times 10^{-16}$\\
	4&$678$&$5730$&$1.1853\times 10^{-10}$&$3.1801\times 10^{-16}$\\
	5&$692$&$5716$&$2.0163\times 10^{-11}$&$8.4063\times 10^{-14}$\\
	6&$700$&$5708$&$9.6533\times 10^{-11}$&$8.2681\times 10^{-11}$\\
	\bottomrule
  \end{tabular}
\caption{Singular values to validate existence of a numerical zero.}
\label{tab:numzero}
\end{table*}
\end{center}

\vspace{-.95cm}
\subsection{Eigenvalue Verification and Maximum Jordan Chain Length}
\label{sec:taxis:eigendecomp:verify0eigenval}
Since the kernel dimension of $A$ (the geometric multiplicity of the zero eigenvalue) is $446$, and there are 699 eigenvalues of magnitude close to zero (as seen in Figure~\ref{fig:realworld_roads_abs_a}),    $699-446=253$ generalized eigenvectors must be determined to complete the Fourier basis.  

As discussed in Section~\ref{sec:AIM:reallworld:sparse}, there is ambiguity in determining whether the 699 eigenvalues of small magnitude represent true zero eigenvalues. Consequently, accurately identifying eigenvalues  of high algebraic multiplicity can be a hard problem -- see, for example,~\cite{zeng_jordanformbkch,zeng_defeigenval,trefethen2005spectra,ruhe1970,golub1976,golub2013matrix}. Furthermore,  these eigenvalues of small magnitude form constellations with centers close to zero, as shown in Figure~\ref{fig:realworld_roads_complexplane_b}. References~\cite{zeng_jordanformbkch} and~\cite{zeng_defeigenval} show that the presence of such constellations may indicate the presence of an eigenvalue of high algebraic multiplicity;~\cite{zeng_jordanformbkch} states that an approximation of this eigenvalue may be the center of the cluster of numerical eigenvalues, while~\cite{zeng_defeigenval} presents an iterative algorithm to approximate a ``pseudoeigenvalue.''  It is unclear from pure inspection\footnote{Considering double-precision floating point (64 bit) and that the number of operations to compute the eigenvalues of an $N\times N$ matrix is $O(N^3)$, the expected precision is on the order of $10^{-6}$ or $10^{-7}$. Numerous eigenvalues in Figures~\ref{fig:realworld_roads_abs_a} and~\ref{fig:realworld_roads_complexplane_b} demonstrate this order of magnitude.} whether the observed constellations are a result of the multiplicity of a zero eigenvalue or are the actual eigenvalues of~$A$. 
For these reasons, it is necessary to verify the existence of a numerical zero eigenvalue before computing the Jordan chains. 
Our method is explained below; see also~\cite{deriasilomar2015}.

To verify a numerical zero eigenvalue, a result from~\cite{ruhe1970} is applied, which states the following:
\begin{result}[\cite{ruhe1970}]
\label{result:ruhe}
Consider a matrix $A\in\mathbb C^{N\times N}$ with singular values $\sigma_1(A)>\dots>\sigma_N(A)$ that is scaled so that~$\sigma_1(A)=1$. Let $m_k=\left| \mathrm{Ker}(A^k)\right|$ denote the dimension of the kernel of~$A^k$. In addition, let~$\alpha$ and~$\delta$ be positive constants; $\delta$ is usually on the order of machine precision and~$\alpha$ is significantly greater than $\delta$. Then~0 is a numerically multiple eigenvalue with respect to~$\alpha$ and~$\delta$ if 
\begin{equation}
\label{eq:ruhe_numzero}
\sigma_{N-m_k} \left(A^k\right) > \alpha >\delta> \sigma_{N-m_k+1}\left(A^k\right),
\end{equation}
for $k=1,2,\dots,h$, where $h$ is the maximum Jordan chain length for the zero eigenvalue.
\end{result}
Since the constants $\alpha$ and $\delta$ have different orders of magnitude, Equation~\eqref{eq:ruhe_numzero} implies that singular value~$\sigma_{N-m_k} (A^k)$ is significantly greater than $\sigma_{N-m_k+1}(A^k)$. 

Result~\ref{result:ruhe} serves two purposes for our application. First, it verifies the existence of a numerical zero eigenvalue. It also implies that a value of $k$ at which Equation~\eqref{eq:ruhe_numzero} fails cannot be the maximum Jordan chain length of the zero eigenvalue. Accordingly, we propose the following method to find the maximum Jordan chain length~$h$: increment the value of $k$ starting from~$k=1$, and let $k=k'$ be the first value of $k$ such that Equation~\eqref{eq:ruhe_numzero} fails. Then the maximum Jordan chain length for eigenvalue zero is $h=k'-1$.

Table~\ref{tab:numzero} is obtained by applying Result~\ref{result:ruhe} to the adjacency matrix~$A$ of the Manhattan road network. The columns of the table correspond to the power~$k$ of~$A$, the dimension $m_k$ of the kernel of~$A^k$, the index~$N-m_k$ of the first singular value of $A^k$ to examine, and the values of the singular values at indices~$N-m_k$ and~$N-m_k+1$. The results in the table are reasonable since the computational machine precision was on the order of~$10^{-16}$, and the first four rows of the table display singular values for which $\delta$ is  on the order of $10^{-15}$ or $10^{-16}$ and the constant $\alpha$ is  larger by 6 to 12 orders of magnitude. Therefore, the inequality~\eqref{eq:ruhe_numzero} holds for the first four rows of Table~\ref{tab:numzero}. The inequality begins to fail at $k=5$; thus, the expected maximum numerical Jordan chain length is no more than~3 or~4.
\subsection{Jordan Chain Computation}
\label{sec:taxis:eigendecomp:jordanchaincompute}
The steps for finding the eigenvectors for eigenvalue zero are described here. This is done to provide ground truth to assess the application of the AIM~\eqref{eq:AIMgft} to real data as shown in Section~\ref{sec:taxis:AIM}.
 
To find the eigenvectors for eigenvalue zero, the null space or kernel~$\mathrm{Ker}(A)$ of $A$ is first computed to find the corresponding eigenvectors. Each of these eigenvectors corresponds to a Jordan block in the Jordan decomposition of $A$, and the Jordan chains with maximum length~$h$ (as determined by applying Result~\ref{result:ruhe} from~\cite{ruhe1970} as shown in Table~\ref{tab:numzero}) are computed by the recurrence equation~\cite{lancaster1985}
\begin{equation}
\label{eq:jordanchain:intaxisec}
A v_k = \lambda v_k + v_{k-1} = v_{k-1},
\end{equation}
where $k=1,\dots,h$ and~$v_0\in \mathrm{Ker}(A)$. If the number of linearly independent proper and generalized eigenvectors equals $N$, we are done; otherwise, the Fourier basis must be extended.

\textbf{Numerical instability.} Numerically stable methods such as the power method, QR algorithms, and Jacobi methods for diagonalizable matrices are applicable methods for eigendecomposition; see~\cite{golub2013matrix} and the references therein for more details. Furthermore, eigendecompositions of large, sparse matrices can be computed with iterative Lanczos or Krylov subspace methods~\cite[Chapter~10]{golub2013matrix},~\cite{bujanovic2011_thesis}.  

On the other hand,  for defective or nearly defective matrices, small perturbations in network structure due to numerical round-off errors can significantly perturb the numerical eigenvalues and eigenvectors; for example, for a defective eigenvalue of $A$ corresponding to a $p$-dimensional Jordan block, $O(\epsilon)$ perturbations in $A$ can result in $O(\epsilon^{1/p})$ perturbations in $\lambda$~\cite{golub2013matrix,wilkinson1965algebraic}. In addition, while computing the Jordan normal form can be numerically unstable~\cite{ruhe1970,ruhe1970_veryill,golub1976,golub2013matrix}, forward stability was shown for the Jordan decomposition algorithm in~\cite{wilkening2007algorithm} and forms the basis for an SVD-based implementation here.  
%
The Jordan chains are generated from the vectors in the kernel of $A^h$, where $h$ is the maximum Jordan chain length; these computations are based on the SVD or the QR decomposition, and so are more stable. Then the Jordan chain can be constructed by direct application of the recurrence equation~\eqref{eq:jordanchain:intaxisec}.  
%



%
\begin{figure*}[tb]
\captionsetup[subfigure]{justification=centerfirst,singlelinecheck=on}
\begin{subfigure}[b]{.53\textwidth}
\includegraphics[width=1\linewidth]{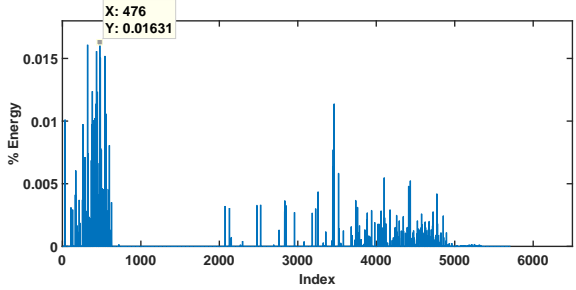}
\caption{Eigenvector spectral components.} \label{fig:percenergy_orig}
\end{subfigure}
\begin{subfigure}[b]{0.53\textwidth}
\includegraphics[width=1\linewidth]{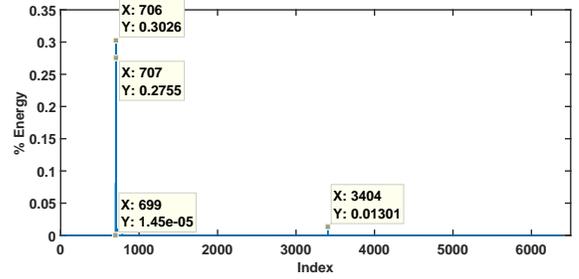}
\caption{Generalized eigenspace spectral components.} \label{fig:percenergy_new}
\end{subfigure}
\hspace{1cm}
\begin{subfigure}[b]{.53\textwidth}
\includegraphics[width=1\linewidth]{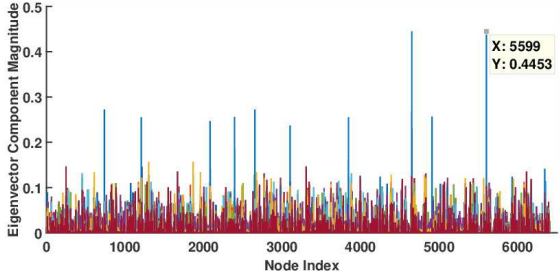}
\caption{Eigenvector component magnitudes.} \label{fig:eigenvecs_orig}
\end{subfigure}
\begin{subfigure}[b]{0.53\textwidth}
\includegraphics[width=1\linewidth]{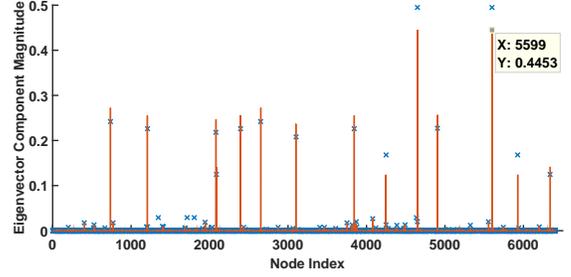}
\caption{Eigenvector component magnitudes.} \label{fig:eigenvecs_new}
\end{subfigure}
\caption[Percentage of energy captured by spectral components and eigenvectors that capture 60\% of signal energy.]{(a) Percentage of energy captured by signal projections onto eigenvectors, including generalized eigenvectors. The horizontal axis corresponds to the eigenvector of the Manhattan road matrix based on the ascending order of the corresponding eigenvalue magnitudes. (b) Percentage of energy captured by signal projections onto generalized eigenspaces. The horizontal axis corresponds to the generalized eigenspace of the Manhattan road matrix based on the ascending order of corresponding eigenvalue magnitudes. The first projection energy onto the generalized eigenspace of $\lambda = 0$ is shown at index 699, corresponding to less than 1\% of the total signal energy. The preceding indices are set to zero to match the indices of (a). The following indices correspond to the same eigenvectors as in (a). The most energy is captured by projections onto the eigenvector at index 706 (30\% of the energy) and  at index 707 (28\% of the energy). (c) Eigenvectors of (a) that capture 60\% of the signal energy. The horizontal axis corresponds to a node (geo-coordinate) in the Manhattan road network. The vertical axis corresponds to the magnitude of the corresponding eigenvector component. There are 84 eigenvectors corresponding to about 60\% of the signal energy, each shown in a different color. (d)  The two eigenvectors in (b) that correspond to 58\% of the total signal energy, one depicted as red lines and the other as blue x's. The axes are identical to those in (c).  The eigenvector components of highest magnitude correspond to the same nodes, or locations in New York City. Furthermore, the locations with highest eigenvector expression match those in (c).}
\label{fig:percenergy}
\end{figure*}
\textbf{Computation details.} A single pass of the Jordan chain algorithm requires about a week to run on a 30-machine cluster of 16 GB RAM/16-core and 8 GB RAM/8-core machines; however, the algorithm does not return a complete set of 253 generalized eigenvectors. To maximize the number of recovered generalized eigenvectors, successive computations were run with different combinations of vectors that span the kernel as starting points for the Jordan chains. After three months of testing,  the best run yielded 250 of the 253 missing Jordan chain vectors; the remaining eigenvectors were computed as in~\eqref{eq:AIMV}. The maximum Jordan chain length is two, which is consistent with Result~\ref{result:ruhe} and Table~\ref{tab:numzero}. As a result, the constructed eigenvector matrix captures a  large part of the adjacency matrix structure. The next section describes our inexact alternative to full computation of the Jordan chains, which provides a significant speedup in the determination of a useful basis set.
\subsection{Generalized Eigenspace Computation}
\label{sec:taxis:eigendecomp:generalizedeigenspacecompute}
In order to compute the generalized eigenspace for the zero eigenvalue, the known eigenvectors are determined as in Section~\ref{sec:taxis:eigendecomp:jordanchaincompute}. Then the eigenvector matrix is computed as~\eqref{eq:AIMV}. This is a five minute computation on a 64-bit machine with 16 cores and 16 GB RAM.
\subsection{Demonstration of the Agile Inexact Method}
\label{sec:taxis:AIM}
In this section,  the original graph Fourier transform, defined as eigenvector projections, is compared to the Agile Inexact Method (AIM)~\eqref{eq:AIMV} for a graph signal representing the four-year average of trips that pass through Manhattan from June through August on Fridays from 9pm to 10pm. 

\begin{figure*}[t]
\captionsetup[subfigure]{justification=centerfirst,singlelinecheck=on}
\begin{subfigure}[b]{0.45\textwidth}
\includegraphics[width=1\linewidth]{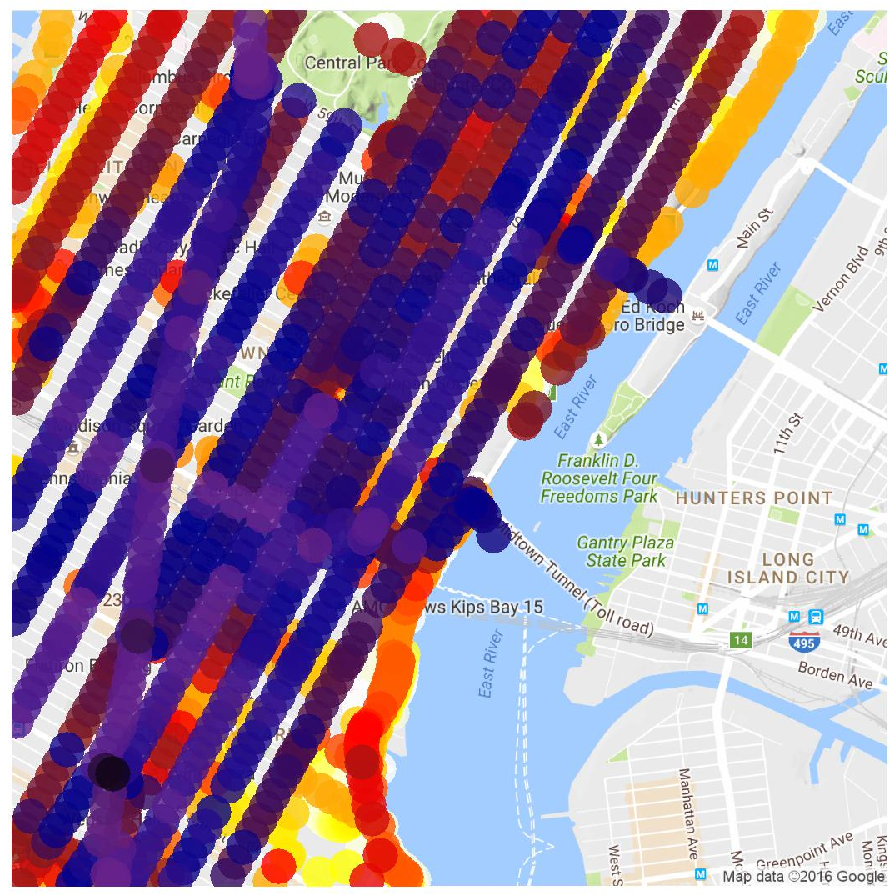}
\caption{Original signal.} \label{fig:map:orig}
\end{subfigure}
\hfill
\begin{subfigure}[b]{.45\textwidth}
\includegraphics[width=1\linewidth]{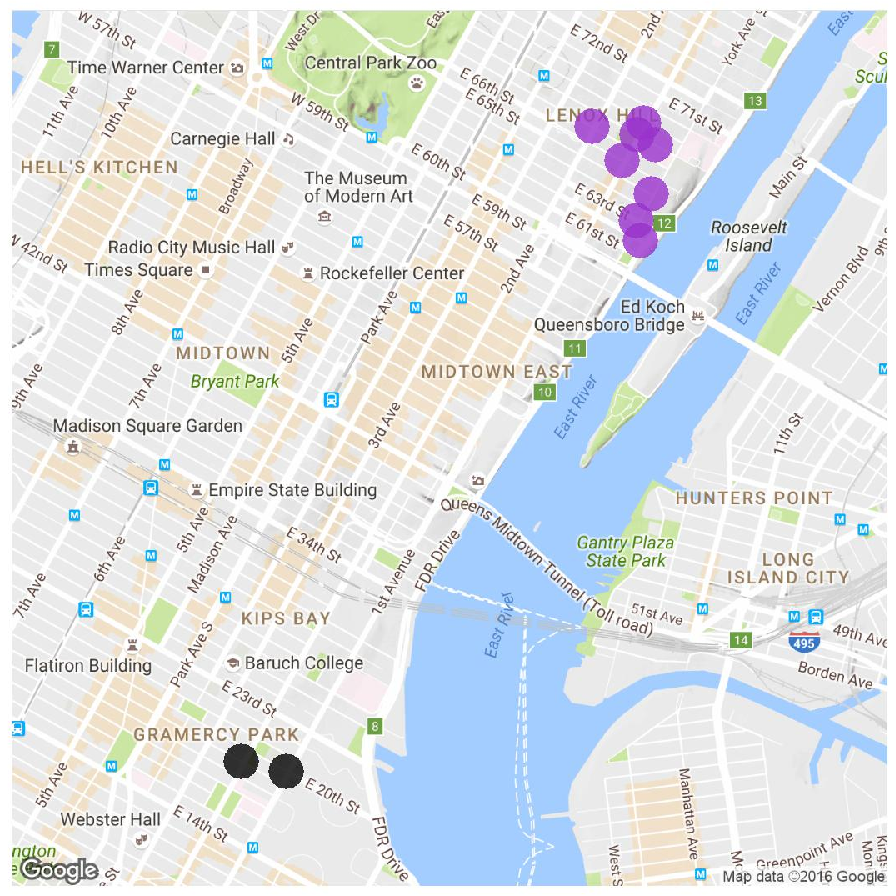}
\caption{Maximum eigenvector components.} \label{fig:map:eigenvec}
\end{subfigure}
\caption{(a) Original signal, representing the June-August four-year average number of trips through each node on the Manhattan road network. Colors denote 699 log bins (white: 0--20, beige to yellow: 20--230, orange: 230--400, red: 400--550, blue: 550--610, purple to black: 610--2,700). (b) Highly expressed eigenvector in Figures~\ref{fig:eigenvecs_orig} and~\ref{fig:eigenvecs_new}, where the maximum components and second maximum components are represented by black and purple dots, respectively. These clusters correspond to locations through which high volumes of taxi trips flow.}
\label{fig:map}
\end{figure*}
Figures~\ref{fig:percenergy_orig} and~\ref{fig:percenergy_new} show the percentage of energy that is contained in the signal projections onto eigenvectors and onto generalized eigenspaces, respectively, using our previously defined energy metric (Section~\ref{sec:AIM:parseval}).  The maximum energy contained in any of the eigenvector projections is 1.6\% as shown in the data point in Figure~\ref{fig:percenergy_orig}. The observed energy dispersal in the eigenvector projections corresponding to $\lambda=0$ is a result of oblique projections. The computed Jordan chain vectors are nearly parallel to the proper eigenvectors (angles less than~1 degree), so the signal projection onto a proper or generalized eigenvector~$v$ of $\lambda=0$ parallel to $\mathbb C^N\backslash \mathrm{span}(v)$ has an augmented magnitude compared to the orthogonal projection. The reader is referred to~\cite{behrens1994signal} for more on oblique projectors.  

In contrast, projecting onto the generalized eigenspaces with the AIM method~\eqref{eq:AIMgft} concentrates energy into two eigenspaces. The data points in Figure~\ref{fig:percenergy_new} at indices 706 and 707 contain 30\% and 28\% of the signal energy, respectively. On the other hand, the energy concentrated on the generalized eigenspace for eigenvalue zero is less than 1\%, as shown by the data point at index~699 in Figure~\ref{fig:percenergy_new}. Thus, the AIM method decreases energy dispersal over the spectral components and reduces the effect of oblique projections shown in Figure~\ref{fig:percenergy_orig}. 

For both methods (the eigenvector projections and the AIM), the spectral components are arranged by the energies they contain, in decreasing order, and those that contain 60\% of the signal energy in the projections have their component magnitudes plotted in Figures~\ref{fig:eigenvecs_orig} and~\ref{fig:eigenvecs_new}. There are 84 such eigenvectors for the projections onto eigenvectors, compared to only two eigenspaces (eigenvectors in this case) that contain the most energy in the generalized eigenspace case; this illustrates that the eigenvector projections disperse more energy than the generalized eigenspace projections.  Figures~\ref{fig:eigenvecs_orig} and~\ref{fig:eigenvecs_new} show that the highly expressed eigenvector components are located at the same nodes for both methods. This demonstrates that the AIM can provide similar results to the original formulation, with a significant acceleration of execution time and less energy dispersal over the spectral components.

The highly expressed components of eigenvectors 706 and 707 in Figure~\ref{fig:eigenvecs_new} are plotted on the Manhattan grid in Figure~\ref{fig:map:eigenvec}. There are two clusters on the east side of Manhattan, one in black in Gramercy Park and one in purple by Lenox Hill. These locations represent sites of significant behavior that suggest concentrations of taxi activity. The corresponding graph signal, shown in Figure~\ref{fig:map:orig}, confirms this. Since this signal represents the average number of taxi trips on Fridays from 9pm to 10pm, one possible explanation for this behavior is the proximity of the expressed locations to restaurants and eateries that are known gathering spots for taxi drivers. Visual comparison of Figure~\ref{fig:map:orig} and~\ref{fig:map:eigenvec} highlights the utility of our method in finding fine-grained behaviors that are otherwise non-obvious from the raw signal. 

This section provides a detailed explanation of a Jordan chain computation for the Manhattan road network. The expensive nature of this computation is a significant drawback for many real-world applications. On the other hand, the inexact computation reveals the same eigenvector expression with computation time on the order of minutes. In addition, the AIM reduces energy dispersal among the spectral components corresponding to the zero eigenvalue.
	
\section{Conclusion}
\label{sec:conclusions}
This paper presents an inexact transform that approximates the spectral projector-based graph Fourier transform defined in~\cite{deriGFT2016}. In Section~\ref{sec:AIM}, the Agile Inexact Method (AIM) for computing the graph Fourier transform is developed for which the generalized  eigenspaces are the spectral components. The generalized Parseval's identity for the inexact method is compared to that of the Jordan subspace projector method in~\cite{deriGFT2016}.  In addition, we show that the total variation-based ordering of spectral components does not change under the inexact computation. The AIM simplifies the eigendecomposition step in real-world, sparse networks, and is associated with a fidelity-runtime trade-off.  Motivated by the equivalence classes discussed in~\cite{deriEquiv2016}, we show that this transform is \emph{agile} and describe the necessary steps for applying our method to real-world data. 

We demonstrate the AIM on a graph signal based on New York City taxi trip data, which yields highly expressed locations that are compared to the ground truth. 
The AIM is shown to reduce the computation time for the eigendecomposition since the Manhattan road network is in the same $\mathscr G$-equivalence class as a graph with an eigenvector matrix that can be computed quickly.  The AIM also reduces energy dispersal among the spectral components compared to oblique eigenvector projections due to the small angles between the Jordan chain vectors and their corresponding eigenvectors.
%
%
%
\bibliographystyle{IEEEbib}
\bibliography{refs}
\end{document}